# Social VR and multi-party holographic communications: Opportunities, Challenges and Impact in the Education and Training Sectors

Mario Montagud*[1,2], Gianluca Cernigliaro[1], Miguel Arevalillo-Herráez[2], Miguel García-Pineda[2], Jaume Segura-Garcia[2], Sergi Fernández[1]

*Corresponding author

[1] i2CAT Foundation, Barcelona (Spain)

[2] Universitat de València (UV), Spain

{mario.montagud, gianluca.cernigliaro, sergi.fernandez}@i2cat.net; {Miguel.Arevalillo, miguel.garcia-pineda, jaume.segura}@uv.es

*Abstract* — Technological advances can bring many benefits to our daily lives, and this includes the education and training sectors. In the last years, online education, teaching and training models are becoming increasingly adopted, in part influenced by major circumstances like the pandemic. The use of videoconferencing tools in such sectors has become fundamental, but recent research has shown their multiple limitations in terms of relevant aspects, like comfort, interaction quality, situational awareness, (co-)presence, etc. This study elaborates on a new communication, interaction and collaboration medium that becomes a promising candidate to overcome such limitations, by adopting immersive technologies: Social Virtual Reality (VR). First, this article provides a comprehensive review of studies having provided initial evidence on (potential) benefits provided by Social VR in relevant use cases related to education, such as online classes, training and co-design activities, virtual conferences and interactive visits to virtual spaces, many of them including comparisons with classical tools like 2D conferencing. Likewise, the potential benefits of integrating realistic and volumetric users' representations to enable multi-party holographic communications in Social VR is also discussed. Next, this article identifies and elaborates on key limitations of existing studies in this field, including both technological and methodological aspects. Finally, it discusses key remaining challenges to be addressed to fully exploit the potential of Social VR in the education sector.

*Keywords* —**Social VR, Virtual Reality, Holographic Communications, VR Education, Immersive Education**

# 1. Introduction.

Information and Communication Technology (ICT) technologies, including software and hardware solutions, have significantly evolved and acquired increasing relevance in the last years, in many sectors of our society. One key example is audiovisual communication technologies, like videoconferencing. Another key example is Virtual Reality (VR) – and immersive technologies, in general - which has been around for the last decades, with a special focus on entertainment, but has acquired wider applicability and adoption due to recent advances in the development and authoring frameworks, content formats, and VR hardware (e.g., cameras, 3D sensors, displays…).

With the arrival of the worldwide pandemic, the adoption of communication, collaboration and consumption tools has been magnified as a response to the lockdown, social distancing and remote working measures. These tools have allowed people to connect with others and reach target audiences, not just for sporadic purposes like in pre-pandemic times, but intensively to keep daily activities and somehow palliate the social isolation. The above-cited examples, videoconferencing and immersive technologies, thus have become key tools for many different sectors of our society, like education, telework and culture, requiring major adaptations and innovations by professionals from all these sectors, but also a mindset change by the whole population as users of such digital tools.

Even acknowledging that digital communication platforms and tools do not reach the levels of face-to-face meetings and physical environments, and still encounter key limitations, their massive usage during these last years has reinforced the wide benefits such solutions provide, like saving time and costs, and contributing to the green deal by reducing the carbon footprint.

Thus, digital communication, collaboration and consumption tools span from just emergence substitute solutions, and have become essential parts of our daily lives. Online and virtual events, activities and experiences have become well established in key sectors like education, training, culture and tourism. Herein, VR has the potential of providing authentic, realistic and immersive experiences, not just for individuals, but also for groups of users. Although Collaborative Virtual Environments (CVE) were devised in the 90s (e.g. [Sto93]), the *Social VR* term has recently emerged and attracted interest from both industry and academia, providing multiple platforms and technological solutions to enable remote users to socially interact, collaborate and/or conduct tasks together [Mon22, Fer22a]. Indeed, Social VR was already anticipated as a new "killer app" that would allow the new generation of VR to break into the mainstream a few years ago [Per16].

Given these new trends and reality, this paper aims to reflect on the opportunities, challenges and potential impact of Social VR in relevant facets of education, including online lessons, meetings, training and (interactive) exploration of virtual spaces. The focus is not only on identifying key benefits compared to the use of traditional 2D videoconferencing platforms but also on discussing the potential benefits that the use of multi-party realistic holographic communications can provide in this domain, as well as on key limitations of the studies conducted so far. Finally, the paper will identify key remaining challenges to be addressed to fully exploit the potential of this new communication and collaboration *medium*, in the path toward scalable, realistic and truly interactive metaverse-like services and experiences, which can open up new opportunities in the education sector.

## 2. Related Work

This section provides a review of state-of-the-art studies providing evidence on the benefits of adopting Social VR solutions in the education sector and related use cases, like training, organization of virtual conferences and events, as well as for virtually visiting and exploring touristic and cultural exhibitions and spaces. Finally, this section identifies related studies identifying and highlighting potential benefits of providing realistic user representations in Social VR, thus enabling multi-party 3D holographic communications within shared virtual environments.

### 2.1. *Social VR for Education*

The arrival of the pandemic implied a wider and more pronounced adoption of digital tools in the educational sector. Examples include [Agu20, Che21]: online educational platforms (e.g., Blackboard, Moodle), communication tools (e.g., Zoom, Teams), social media (e.g., TikTok, Linkedin, Facebook), asynchronous videos, and synchronous class sessions (live).

Recent scientific studies have emphasized and demonstrated many potential benefits that multi-party videoconferencing solutions can provide in the education sector, especially when leveraging interactive and collaborative features supported by such solutions (e.g. [Bor21]). As discussed earlier, the interest in such platforms, as well as in exploiting their capabilities, has been magnified with the arrival of the pandemic (e.g. [Che21], [Hag22]).

In addition, the use of immersive technologies, virtual environments and Social VR platforms has also started to attract higher interest in this domain (e.g., [Rad20]). Many research studies have shown that several affordances of VR, including interactivity and immersion, provide positive outcomes in the education sector [Rad20], increasing presence, motivation, engagement and learning abilities [Yos21].

The study in [Gre13] reviewed the use of virtual environments at 19 surveyed educational institutions, for many different purposes like role play activities, virtual tours, staff or faculty development, etc., using primarily standard desktop interfaces. The study in [Agu20] states that the use of immersive technologies in education can positively influence the attitudes and behaviours of both individuals and groups, and can also contribute to self-directed learning. The study in [McG21] analyzes the adoption of VR technologies in the education sector, identifying where more educational implementation and research need to be done, and providing a perspective on future possibilities focusing on current affordances.

In this context, many studies have focused on comparing traditional videoconferencing platforms with Social VR platforms, as well as comparing (2D) desktop and (3D) VR headset viewing conditions, in different educational scenarios. The study in [Ste20] preliminarily compared group meetings in a video conferencing platform, *Zoom*[1], and in a web-based Social VR platform, *Mozilla Hubs*[2], using both desktop (2D) and headset (3D) viewing conditions. Higher social presence levels were reported for VR viewing in Mozilla Hubs than for traditional conferencing. However, usability levels were lower for Social VR. The study in [Mur16] found that 3D VR viewing provided higher spatial learning levels compared to 2D desktop viewing when using a developed ad-hoc virtual environment. The study in [Obe19] showed higher enjoyment and learning quality levels for 3D VR viewing than for 2D desktop viewing, also when using a developed ad-hoc virtual environment focused on a serious game context.

Similarly, the study in [Che21] compared a traditional videoconferencing platform, *Teams*[3], and a Social VR platform, Mozilla Hubs, in 2D desktop viewing mode, in terms of sense of presence and usability for online (university) lectures. That study suggests that even though Social VR systems reduce usability (i.e., they are less easy to use, probably due to lack of familiarity), they increase presence when compared to traditional videoconferencing systems (even when using 2D screens and avatar-based representations), also concluding that both types of tools seem to be effective for conducting online lessons. In addition, that study does not only reflect on the potential of Social VR for the educational sector but also highlights the need for further research on determining and exploiting its benefits.

The study in [Yos21] explored student experiences for remote instruction (University level) using Social VR, in particular Mozilla Hubs platform, with students attending classes from home for 7 weeks (Figure 1). That study focused on evaluating experiences (within-subject methodology) when: (i) attending / viewing remote lectures with VR headsets; (ii) attending / viewing remote lectures with desktop displays; (iii) making presentations with VR headsets. The evaluations considered social presence, simulator sickness factors, communication methods, avatar and application features, and tradeoffs with other remote approaches. The study also aimed at gaining insights from having participated in such a VR-based remote class running for several weeks. Overall, results from [Yos21] suggest that Social VR platforms provide a promising alternative to other remote class approaches, despite the discomfort in some cases and sporadic technical issues. 3D VR headset viewing and presenting produced a higher presence than 2D desktop viewing, but had a less-clear impact on social presence and other overall experience measures. The use of 3D VR headsets for making and attending presentations resulted in an attentional allocation over the use of 2D desktop screens. In addition, a trend toward higher perceived emotional and behavioural interdependence was identified once using 3D VR headsets over 2D desktop screens for making presentations. Sickness symptoms occasionally appeared but were primarily related to general discomfort, rather than to motion. Desktop viewing was found to be an adequate alternative for those students having felt uncomfortable and/or simulation/VR sickness, or those having experienced technical issues. Indeed, that later group of students reported a preference on adopting mixed methods between headset and desktop viewing, as a feature to be implemented. Students also reported feeling less nervous giving presentations as avatars than they would when using conventional video.

In line with the insights from [Yos21], the study in [Ryu21] compared desktop, VR and a mixed viewing mode for lectures in Social VR, using Mozilla Hubs platform. The mixed mode used a custom fixture to simplify switching between desktop and VR viewing modes, and it was well received by participants given its ability to combine the best of each mode.

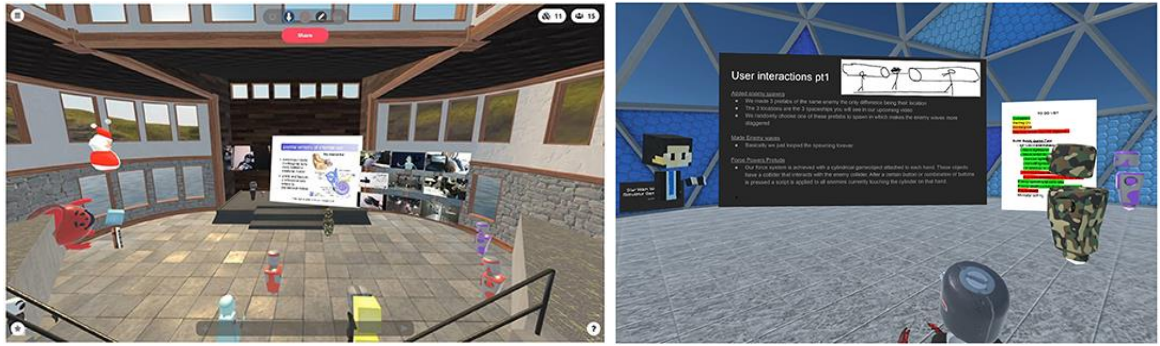

*Figure 1.  Lectures and presentations in Mozilla Hubs, as recreated in [Yos21]*

## 2.2. *Social VR for Training*

VR products and experiences have often attracted interest in the industry and training sectors, given the potential of recreating shared interactive and collaborative environments. For instance, trainees or even professionals can be exposed to realistic scenarios that may involve high costs or risks in real life, allowing them to gain experience before actually undertaking the associated real situations. While these virtual experiences were initially conceived for individual usage, interactive and multiuser experiences are becoming increasingly popular with the latest advances and trends.

The study in [Ber17] provides a review of a wide set of industrial VR applications, from which a collaborative design use case is highlighted, allowing engineering, design and marketing teams to get together for optimizing specific products. The study in [Ble20] highlights the potential of VR technologies in effectively supporting interactive conversations and collaborations between remote users, although it reflects on the need for a collaborative design framework addressing key factors and aspects, such as: (i) variability in social immersion; (ii) user and conversational roles; and (iii) effective shared (and spatial) references. That study also identified practical issues that can become social barriers, like not being able to see what other users are looking at in a virtual environment, thus reducing the illusion of sharing an experience.

The study in [Kho21] provides a longitudinal and exploratory study on individual workload, presence, and emotional recognition in collaborative multi-party virtual environments. The obtained results show that although the reported presence and workload levels did not vary over time, the adaptation to VR (in terms of interaction with partners and execution of tasks) significantly increased over time, and the reported co-presence levels were influenced by the task at hand. That study also discusses design implications and suggests future directions for designers and researchers in the field, with a particular focus on how to effectively enable meaningful social interactions and collaboration in VR.

The work in [Mei21] adopted a Social VR platform to explore the appropriateness of this medium to solve co-design tasks, like allowing pastry chefs and clients to get together for proposing, refining, customizing and validating cake designs, and exploring them interactively in 3D from any viewpoint (Figure 2). Although limited to a pilot experience, both groups of participants were enthusiastic about the potential and flexibility of this medium for collaborative design and prototyping.

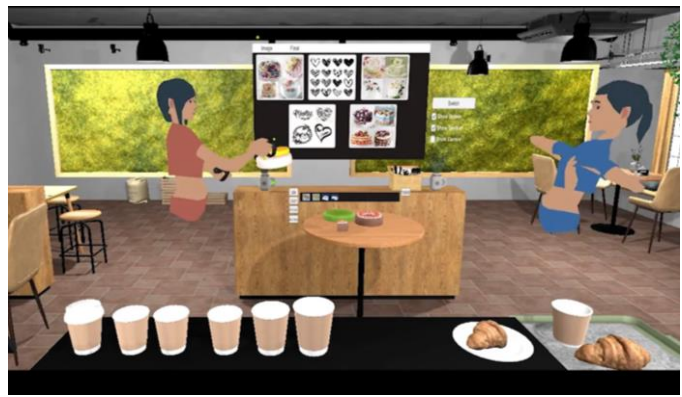

*Figure 2. Social VR for co-design of cakes, as explored in [Mei21]*

In this line, the study in [Jon22] reflects on the potential of Social VR for rapid prototyping and co-design by remote persons, anytime and anywhere, minimizing costs (e.g., food, materials, hardware…), allowing replicability, reversible step-by-step works that can be saved at any point, interactive 3D inspection from any viewpoint, and a high degree of customization. However, that study also highlights learning curves and usability issues, especially once involving novice participants, and once performing user tests using VR hardware. In addition, that study also emphasizes the need for Social VR platforms for reproducing non-verbal communications and gestures to recreate realistic and natural gestures, movements and actions, as they become fundamental to enable effective collaborative and rich understanding in many training disciplines.

2.3. *Social VR for Conferences and Events*

Social VR has also become a promising medium to hold virtual conferences and events, attracting interest due to its potential to contribute to sustainability (i.e., reducing the environmental impact due to travel avoidance) and accessibility, and to provide an enhanced experience compared to traditional conferencing platforms (e.g. [Le20], [Che21]). The idea of using VR environments for holding online conferences is not new. In 2010, the *web.alive*[4] tool from Avaya was adopted to allow mixed participation in the 3rd International Workshop on Massively Multiuser Virtual Environments (MMVE 2010) [Shi12] (see Figure 3, capture on the left). Remote participants highly valued the option to virtually participate in the event, especially due to their unavailability to attend in person. In 2011, researchers from IBM examined the potential of organizing an entirely virtual corporative conference in an avatar-based 3D virtual environment using Second Life[5] [Eri11] (see Figure 3, capture on the right). The conference was reasonably successful, although technical issues related to the platform set-up, performance and stability were quite frequent.

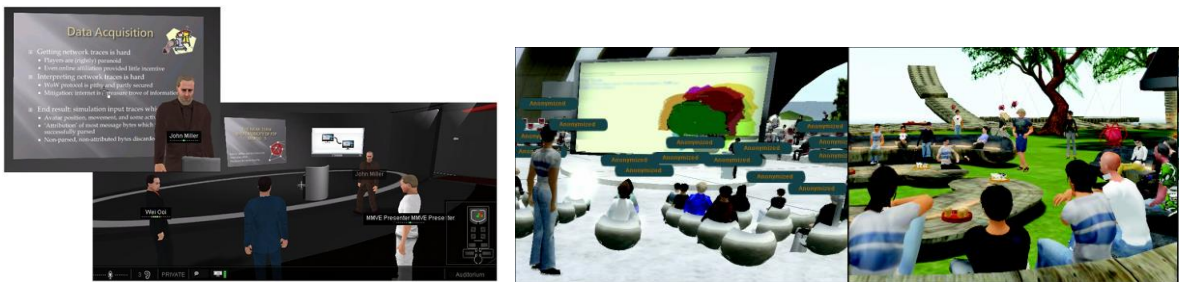

*Figure 3.  Social VR for virtual conferences: (left) event recreated in [Shi12]; (right) event recreated in [Eri12]*

The study in [Neu16] tested the use of telepresence robots for enabling remote attendance at the ACM International Joint Conference on Pervasive and Ubiquitous Computing 2014 (Ubicomp'14). Attendees with accessibility issues felt especially empowered because they were able to experience more of the conference than they were used to. Additionally, participants stated having felt more immersed than when using traditional video conferencing tools.

The study in [Cam20] compared the use of VR environments, using HTC Vive headset, and the use of a traditional video conferencing software, *Skype for Business*[6], for conducting online meetings in a corporative context. VR meetings resulted in increased focus and engagement, and female participants reported not feeling judged based on their appearance, which suggests that Social VR can contribute to gender equality.

The study in [Le20] reports on insights from having adopted a Social VR platform, Mozilla Hubs, for enabling a shared virtual space for both co-watching the conference talks remotely and virtually attending poster sessions at ACM User Interface Software and Technology conference 2019 (UIST'19). Participants were allowed to use their preferred consumption device. The obtained results reflect that, although Social VR does not reach the benefits of physically attending a conference, it provides a reasonable and successful alternative to accomplish key goals behind attending conferences, like learning from talks, meeting new people, and discussing ideas and/or opportunities with potential collaborators. Participants in general felt involved and immersed, and technical issues were not so common. That study also reflects on the inherent advantages of remotely attending conferences (e.g., time and cost saving, democracy, accessibility, scalability and replicability…), and on the remaining opportunities to fully exploit the unique features of Social VR platforms to enrich the attendees' experience, like environment and avatar personalization, spatial audio, and multi-modal interaction (e.g., shared presentations, teleportation, laser pointing, taking selfies, integration of widely adopted chat tools like Discord…). Finally, the study in [Le20] also reflects on the benefits of having adopted a web-based cross-platform Social VR platform to provide interoperability and stability, and provides recommendations on how to ensure certain levels of scalability, like setting up parallel rooms for each session/activity and/or a lobby room connected to adjacent rooms in a ring/star topology

---

[4] Web.alive: https://support.avaya.com/products/P0942/avaya-webalive, Last access in October 2022
[5] Second Life: https://secondlife.com, Last access in October 2022
[6] Skype for Business: https://www.skype.com/en/business/, Last access in October 2022

## 2.4. *(Social) VR for Culture and Tourism*

The use of (Social) VR tools for cultural and touristic visits has also attracted attention and has shown to provide relevant benefits, including business use cases. Relevant use cases in this context include virtual visits to museums and exhibitions (e.g., [Mon20a, Hag20]) and to current and past cultural heritage (e.g. [Mon20b, Loa20]), among others.

The study in [Bor18] analyzed virtual field trips within networked VR environments for small classes, driven by a teacher. Benefits in terms of presence and social presence were identified, encountering no major simulation sickness and external distraction factors.

The study in [Tat21] compared presence and information retention levels when visiting a real museum and a replica of it in virtual reality. The obtained results show that virtual visits provided similar presence, learning and knowledge retention levels as real visits.

The study in [Hag22] compared the social presence levels associated with a virtual guided tour to a theatre when using a traditional videoconferencing platform, Microsoft Teams, and a Social VR platform, Mozilla Hubs (Figure 4). When using Microsoft Teams, a guide entered the virtual environment and shared the screen with the virtual visitors. The obtained results revealed minor differences between the two tested conditions, which can be attributed to the adoption of 2D (desktop and mobile) viewing, as well as to some encountered usability issues and lack of interactivity features in the VR condition. However, that study reflects on the potential of immersive mediums to support these experiences and provides recommendations on how to effectively enable them.

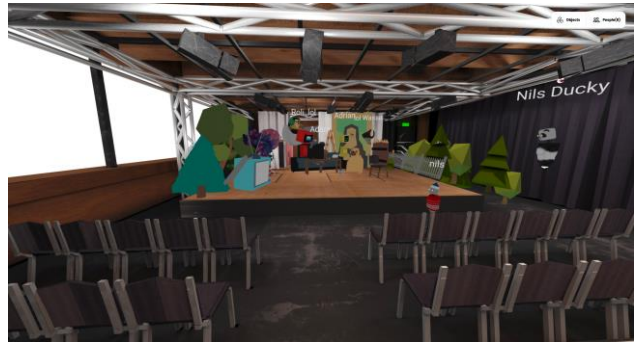

*Figure 4.  Virtual Visit to a Theater using Social VR, as explored in [Hag22]*

The insights and conclusions from all these previous studies suggest that (Social) VR becomes a promising tool for different facets of (remote) education, including attendance at lectures and conferences, training activities and (interactive) exploration of spaces of interest.

### 2.5. *Potential Benefits of Realistic User Representations in Social VR*

Most Social VR platforms rely on the use of synthetic, either cartoon-like or human-like, avatars for representing the users [Mon22, Fer22b]. This has been identified as a barrier for use cases in which a serious and professional appearance is required [Lee21], and can also have an impact in the education sector, as discussed next.

The study in [Yos21] reflected on the importance given by students to the teacher's avatar when performing online lessons using a Social VR platform, as full body tracking and mouth movements were considered key aspects to provide a rich comprehension and communication.

Many other studies have also identified the importance of embodiment for providing immersive experiences (e.g. [Kil12]). The study in [Hei17] compared different types of avatar appearances, and it was found that motion-controlled avatars, even if only having head and hands visible, produced an increased feeling of co-presence and behavioural interdependence over other types of avatars that did not resemble movements and gestures. The study in [Rot16] analyzed whether realistic body movements of full-body avatars could compensate for missing facial expressions and eye gaze cues. The obtained results indicated that social interactions tend to be impeded with non-realistic avatars, but the absence of important behavioural cues, such as gaze and facial expression, can be partially compensated by realistic body movements. The study in [Smi18] compared the quality of audio-visual communication between two users completing a task, under three conditions: (1) face-to-face; (2) ad-hoc Social VR environment, with embodied avatars that have an eyebrow ridge and nose, but no other facial features; and (3) ad-hoc Social VR environment, without visible avatars but only virtual hands. The obtained results revealed that embodied avatars provide a high level of social presence, with conversation patterns that are very similar to face-to-face interaction. Similarly, the study in [Her19] compared the task completion effectiveness when operating machinery in an ad-hoc VR environment. The obtained results revealed that it was easier for participants to complete the tasks when they were guided by an avatar than when they were guided by a disembodied voice.

In addition, many other studies investigated the impact of virtual representation realism on the user experience. The study in [Lat17] explored the effect of avatar realism on self-embodiment and social interactions in Social VR. Realistic avatars were rated significantly more human-like and evoked a stronger acceptance in terms of virtual body ownership. Similarly, the study in [Wal18] found that personalized avatars significantly increase the sense of body ownership, presence and dominance compared to other two types of generic (hand-modelled and scanned) avatars being explored. The study in [Mon22] showed that the availability of realistic and volumetric user representations (i.e., 3D holograms) in Social VR, even when provided by low-cost volumetric video capture setups, provides enhanced presence, social connectedness and togetherness than the use of avatar-based representations, and similar levels than face-to-face settings, in shared video watching scenarios. These are preliminary, but very promising results, especially given the room for improvement that the state-of-the-art volumetric video capture and processing technologies have. In this line, the study in [Fer22a] proved that state-of-the-art technology for holographic communications, despite their current limitations, already enables effective collaborative and interactive tasks in multi-party virtual meetings (Figure 5).

Thus, the studies reviewed in this subsection suggest that the benefits provided by Social VR platforms in the studies reviewed in the previous subsections may be augmented if realistic user representations can be provided, regardless of the target use case.

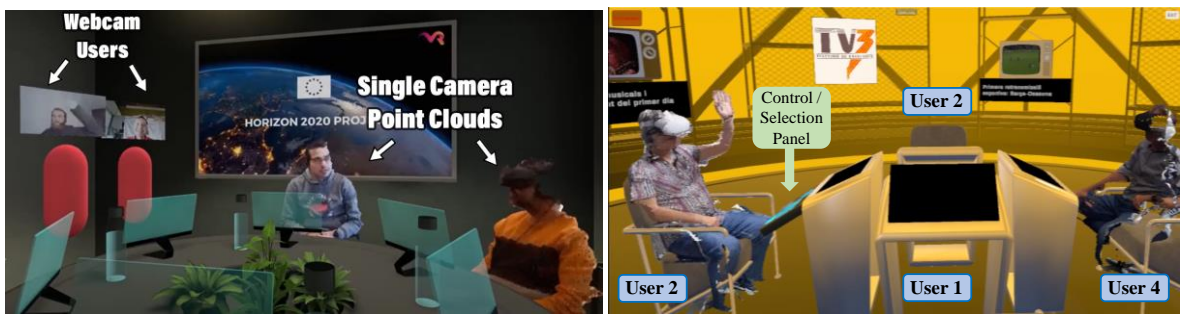

*Figure 5. Examples of Social VR scenarios with holographic communications, as enabled in [Fer22a]*

## 3. Limitations of Existing Studies

Despite the wide sample of studies that have reviewed and investigated the potential applicability and benefits of Social VR in the education and related sectors, key gaps and the need for extra research can still be identified. This subsection elaborates on a selection of the identified limitations.

### 3.1. *Avatar-based Social VR platforms*

Most of the studies and experiments conducted so far have been based on the use of avatar-based Social VR platforms, like Mozilla Hubs. As discussed in Section 2.5, avatar-based representations, even being customizable, typically lack realistic and natural expressivity, gestures and movements, and do not accurately resemble the associated real persons. In some cases, these inconveniences can be tolerated, but in others they can result in a lack of confidence and comfort. In addition, they can result in lower levels of (co-)presence, immersion and social connectedness [Mon22]. Therefore, the necessity for further exploring the adoption of volumetric and realistic user representations in Social VR emerges, as well as the need for determining the associated benefits in the use cases of interest.

### 3.2. *Limited Interactivity and Lack of Realistic Digital Twins*

On the one hand, key strong aspects of VR environments are the ability to integrate rich multi-modal interaction features (e.g., teleporting, interaction and manipulation of 3D objects, laser pointing…) and stimuli (e.g., heterogeneous media formats, multi-sensory effects like haptic feedback and scents…). On the other hand, the advances in media formats, 3D scanning and modelling solutions, together with the availability of powerful processing and display hardware, enable a high-quality and (ultra-)realistic recreation of existing, recovered and/or ideated digital spaces with moderated efforts.

However, the conducted experiments and studies so far have been mostly based on the adoption of simple VR environments. Although such types of environments can serve the purposes of the target studies (e.g., attending and giving lectures, watching a shared video/presentation, having a meeting…), the availability of highly interactive and realistic virtual spaces has the potential of enhancing the user experience in different aspects, but also to positively contribute to learning and training, e.g. through interactively exploring cultural heritage spaces and assets, by manipulating and interacting with realistic reconstructions of installations and machines, etc. The associated benefits of providing these (replicable and adaptable) scenarios in the education and training sectors seem beyond doubt, but it needs to be explored and determined using scientific methods.

### 3.3. *Limited scope and test conditions*

Some limitations regarding the test conditions and scope of the state-of-the-art experiments/studies can also be identified, without meaning to be exhaustive.

First, although various studies have put efforts into comparing 2D desktop and 3D VR viewing conditions in Social VR, conclusive results are not yet available and further research is needed. In this line, reported experiences of using VR headsets and controllers for attending virtual classes/meetings or giving presentations are still very scarce, especially when it comes to distributed settings in open environments (e.g., participants from their homes) [Rad20].

Second, most of the studies conducted so far, except for full conferences (e.g., [Le20]) and courses (e.g., [Yos21]), have a limited duration for experimentation. In addition, the few existing experiments on interactive and collaborative scenarios have been limited to a few participants. Longer-term and larger-scale experiments and pilots are necessary to accurately determine the impact and implications of adopting Social VR in the education and training sectors.

Third, most studies reported so far have been focused on university and post-graduate levels (e.g., scientific conferences). However, Social VR and holographic communications can also potentially provide wide benefits at other educational levels, like elementary/secondary/high school and vocational training. Further research is needed to explore and determine the benefits in all these sectors and disciplines, in close collaboration with associated instructors, pedagogical experts and entities.

### 3.4. *User Experience (UX) Evaluation*

Previous studies on adopting Social VR for the education and related sectors have considered a wide variety of User Experience (UX) and learning factors, including presence, immersion, comprehension and information acquisition levels, engagement and motivation, usability, etc. On the one hand, while the impact on presence and place illusion (i.e., the feeling of "*being there*") has been commonly explored, plausibility effects (i.e., the illusion that the scenario being provided or the experience to which being exposed is occurring) [Sla12] have much been less explored in this domain. On the other hand, collaborative and shared experiences are central to many forms of educational pedagogy [Le18]. However, most studies have been focused on analyzing the impact on presence, but much less on co-presence and togetherness. The study in [McG21] reflects on the need for VR tools and platforms to enable natural and rich interaction, socialization and collaboration, thus providing co-presence, as humans are social by nature and typically experience learning and training activities in groups, or with others. Co-presence allows natural social interaction, where users can move around, see, hear, and in some cases feel each other. In the context of this study, co-presence can be associated to a virtual class, where students virtual meet and naturally interact as if they were together in a real classroom. This also means that users can interact with objects together and socially, writing on the same board, seeing each other's work, and even picking up and engaging with digital objects together.

Thus, the potential of Social VR to effectively enable co-presence and conduct tasks and activities together needs to be further explored.

### 3.5. *Comparison with baseline and alternative test conditions*

The study in [Web16] highlights that previous research on VR education was mostly focused on determining VR-related experience factors, such as immersion, presence, engagement, as well as usability, but less on exploring the levels of knowledge acquisition compared to traditional methods. The recent literature review in [Ham21] states that, although previous studies have reported on key benefits when using VR technologies in the education sector (e.g., in terms of procedural and affective skills), it is necessary to perform thorough comparisons with traditional methods and technologies to determine whether Social VR can be accepted as a reliable and beneficial pedagogical method.

In addition, despite that few studies have considered baseline conditions in their test conditions, they are mostly focused on comparing traditional 2D video conferencing to Social VR platforms (e.g. [Ste20, Che21, Hag22]). On the one hand, the test conditions to be presented to the experiment participants need to be accurately prepared, so that the comparison can be meaningful and derive valuable insights and precise results. Large deviations between test conditions would need to be avoided as much as possible, de-limiting or having under control the key test variables, e.g. in terms of familiarity, stimuli, interaction modalities, etc. On the other hand, comparisons between physical situations and virtual situations, ideally with advanced and interactive scenarios, would need to be conducted to accurately determine the pros and cons of Social VR over real baseline scenarios.

## 4. Remaining Challenges

Strongly aligned with the identified and discussed limitations, remaining challenges to effectively and widely adopting Social VR and holographic communications in the education and related sectors can be identified, which are briefly discussed next.

### 4.1. *Multi-modal interaction and Realistic Digital Twins*

Despite major advances in VR development frameworks, 3D scanning, modelling and reconstruction, as well as authoring and editing tools and on the integration of Internet-of-Things (IoT) data, the creation of highly realistic and interactive virtual environments and Digital Twins of real environments is not yet easy. It typically requires the intervention of expert personnel and considerable (time and cost) efforts, and cross-platform and cross-device compatibility issues still apply. The availability of easier-to-use VR creation tools, larger and more diverse asset repositories, *plug&play* hardware, and guidelines would significantly help toward the creation and offering of highly interactive and realistic virtual spaces, accurately recreating what was planned.

### 4.2. *Realistic Holographic Communications*

As discussed in Section 3.1, the availability of 3D holographic representations can potentially provide key benefits compared to avatar-based representations. Even though recent studies have proved that state-of-the-art technology, using off-the-shelf and low-cost equipment, is already able to effectively offer such multiuser holographic services [Fer22], further work is needed to mainly: (i) improve the visual resolution; (ii) reduce costs (hardware, bandwidth, processing…); (iii) enable *plug&play* solutions; (iv) address interoperability and scalability issues (see next subsection); etc.

### 4.3. *Interoperability and Scalability*

On the one hand, the VR ecosystem suffers from fragmentation. Applications and tools developed for one platform and/or device may not (completely) work for other platforms and/or devices. In addition, VR displays significantly differ regarding their capabilities and/or interaction modalities. Thus, interoperability and compatibility remain still a barrier toward achieving large-scale adoption of Social VR platforms, especially when it comes to the integration of 3D holographic representations and usage of VR consumption devices. Of course, this also applies when it comes to the concurrent usage of different VR consumption devices, and of non-VR devices, for the same application, and in the same multiuser sessions.

On the other hand, Social VR platforms still encounter scalability challenges. Different solutions can be adopted to somehow overcome this. First, in the case of events/conferences, VR rooms can be replicated so virtual attendees can be strategically directed to the available rooms. Typically, a central lobby room can be used to connect other rooms in a star or ring topology (e.g., [Le20]). Second, network-based media processing features can be leveraged, e.g. to reduce bandwidth and processing capabilities at the client side (e.g., [Cer20], [Gun21]). Emerging cloud continuum and (5G and toward 6G) smart network strategies can also contribute to overcoming interoperability and scalability. Third, the virtual sessions can be 2D broadcasted, as it is done in e.g. the gaming community using platforms like Youtube or Twitch. This last solution does not offer fully immersive experiences, but still allows participation or having access in/to them.

All these interoperability and scalability issues can potentially bring service continuity and performance implications, and thus are key technical challenges to be addressed or at least have under control.

Finally, an extra, yet related, aspect is the availability of appropriate VR equipment to be able to run large-scale and long-term pilots/experiences in distributed scenarios.

### 4.4. *Usability and Accessibility*

VR experiences and devices have not yet become mainstream, so people are not yet used to them. In addition, VR usage requires specific competencies and skills, which require a learning curve. The time and competency requirements somehow have had an impact on the adoption of immersive environments in education [Car10]. Without mastery of the VR tools and devices, errors may occur, leading to loss of information, learning effects and sub-optimal experiences for students and teachers. For instance, research has shown that many errors and/or faults occurring in VR usage are related to the user's ability to effectively use the VR equipment rather than to the equipment and/or software itself [Gre15].

While usability and homogeneity issues are being overcome, the use of tutorials, step-by-step walk-through videos and initial simple environments to get familiar with the tool/experience are becoming commonplace in the VR ecosystem (e.g. [Le20, Che21]).

In addition, most traditional media services are provided with accessibility solutions (e.g., accessible user interfaces and interaction modalities, presentation of access services like subtitling, audio description and sign language interpreting…). However, that is not yet the case for immersive services (e.g., [Mon20a, Mon20c]), for which accessibility-related standards, guidelines and/or best practices recommendations are still in their infancy. This fact results in a barrier once targeting large-scale and open pilots/services based on VR technology and needs to be addressed shortly.

### 4.5. *Comfort, Ethics & Privacy*

Comfort, safety, privacy and ethics are key issues to be guaranteed in/by (Social) VR services. Harassment, trolling, as well as incorrect and abusive activity and behaviour, are not easy to detect in (Social) VR, despite that they are significant problems that need to be avoided and penalized [Jon22]. For instance, malicious usage, non-respectful behaviour (e.g., moving too close to other users' virtual space or resources [Sun21]), abuse of anonymity (leveraging the avatar-based representation), and even racist and sexist behaviour, have been reported in previous studies on Social VR [Jon22]. It thus becomes essential to provide safe environments in which comfort, ethics and privacy can be guaranteed.

### 4.6. *Evaluation Methodologies*

As reported in Sections 3.4 and 3.5, several gaps and limitations regarding evaluation metrics and methodologies can be identified from the state-of-the-art review. Often, these types of limitations are derived from a lack of standards, guidelines and best practices recommendations on how to conduct evaluations for Social VR and/or VR education.

State-of-the-art studies have widely adopted established, validated and even standard recommendations, metrics and questionnaires from related VR fields, based on their associated context and targeted goals. These include presence questionnaires, usability questionnaires, task load index questionnaires, recommendations on durations to induce adequate immersion while avoiding simulation sickness and other related symptoms, etc. In addition, recent studies have proposed and validated methodologies for the evaluation of Social VR [Mon22]. However, there is currently neither a reference nor standard evaluation methodology for such a medium, which would allow maximizing (the impact of) insights of related studies and even comparing results across related studies. Such new evaluation methodologies would need to include procedural aspects throughout the lifetime of the study, key metrics to consider, and resources to collect them (e.g., questionnaires, tools…). In addition, given the relevance of the use case(s) being discussed in the paper, these methodologies would need to encompass key educational and pedagogical factors, with the collaboration of experts in these domains.

## 5. Conclusions & Future Work

Technological advances have the potential of making our lives easier and better. In this context, Social VR technologies, initially conceived for entertainment use cases, are attracting interest and experimental adoption in various sectors, like education, tourism, and for holding different types of meetings and events [Mon22]. This increasing interest in Social VR has been fueled by the arrival of the pandemic, and by pioneering studies reflecting on key advantages provided when compared to the usage of traditional 2D conferencing tools, mainly in terms of plausibility, (co-)presence and quality of interaction.

This article has surveyed and categorized existing studies on exploring the potential impact and benefits of adopting Social VR, as an emerging communication and collaboration medium, in different scenarios related to education. This review has provided a general picture of the latest advances and offered services, as well as insights and results from associated pilot actions and evaluations. It has also identified key limitations of existing studies and remaining challenges to be addressed to maximize the potential of Social VR in education and other related key sectors. Of course, these potential and derived benefits will not be restricted to the agents involved in (Social) VR, like content producers, content providers and developers, but also to all agents involved in the education sector, like professors, students, family members, etc. Social VR already brings a wide set of novel and rich opportunities, and these are expected to significantly increase in the near future. It should not be conceived as a substitute medium for in-person activities, but instead, as a supplement to traditional educational spaces, events and experiences, augmenting the possibilities and resources from/in physical scenarios, bringing sustainability, accessibility and cost benefits. Although unceasingly evolving day by day, technological advances are (almost) there, and now it is the time to leverage them to fully exploit new possibilities in key sectors of our society, like virtual education and training. This article thus can serve the interested audience to get familiar with the advances made and to be made in this domain, where both relevant research and exploitation opportunities and necessities exist.

## ACKNOWLEDGMENT

TBC